# Existence Proof of nonThermal Vacuum Radiation from Acceleration*

By John Michael Williams

jwill@AstraGate.net
**Markanix Co.**
P. O. Box 2697
Redwood City, CA 94064

2002 January 31

## Abstract

A proof is developed from first principles, independent of general relativity and of thermodynamics, that there exists a threshold acceleration above which radiation (real particle creation) from the vacuum must occur. The radiation is not expected to follow a Planckian distribution.

## Introduction

We use the generic term, "vacuum-acceleration radiation", to refer to Hawking-Unruh radiation or any other radiation to be shown caused solely by high acceleration in vacuum. Radiation caused by acceleration of charge (synchrotron radiation, for example) is not meant to be included. We wish to prove that real particles of some arbitrary kind will be pulled from the vacuum at sufficiently high acceleration of any arbitrary (massive) particle. Our proof is not limited to soft photons, nor to thermodynamic considerations. However, we assume special relativity to the extent of the Unruh-Davies derivation in [1].

The Einstein equivalence principle implies that some of the radiation near the horizon of a black hole might be vacuum-acceleration radiation, if such radiation existed; this effect of a black hole never yet has been observed astrophysically. Chen and Tajima [2] have proposed a mechanism by which Hawking-Unruh radiation caused by acceleration might be studied in the laboratory. Their approach depended upon quantum field theory and the curved space-time of general relativity.

Belinsky [3] has raised the question that radiation by the Hawking-Unruh mechanism theoretically might not be possible. Rabinowitz [4] has proposed that black-hole radiation might occur because of the gravitational field of a nearby body, which would facilitate quantum tunneling by a mechanism similar to field emission. Parikh & Wilczek [5] have shown that Hawking-Unruh radiation might be viewed as a special case of quantum tunneling through a potential barrier. Their approach is similar to that of Rabinowitz, except that they treat a spherical shell originating from the black hole as the second body.

We assume no theoretical limit on acceleration at least up to the acceleration Einstein-equivalent to that at the horizon of a black hole. How big a black hole, remains open: It seems obvious that any black hole presently existing must have a horizon with acceleration from gravity below the threshold needed to produce vacuum-acceleration particles at a significant rate.

# Thermal Distribution in High Acceleration

Planck's distribution describing a black body (thermal) radiator may be written,

$$I(\mathbf{n}, T) = \frac{2\mathbf{p}}{c^2} h\mathbf{n} \frac{\mathbf{n}^2}{e^{h\mathbf{n}/(k_B T)} - 1}, \qquad (1)$$

in which emission frequency $\mathbf{n}$ (uniform scale) and temperature $T$ determine the radiant spectral density $I$, with $h$ Planck's constant and $k_B$ Boltzmann's constant. The alternate form, in which a uniform wavelength scale is assumed, replaces the $\mathbf{n}^2$ in the numerator with $\mathbf{n}^4$ but has no important effect on the result below.

Several computations reviewed in [7] arrive at the conclusion that radiation from a black hole, or because of other acceleration, will be thermal, in a certain sense. For example, the spectrum from a body in constant acceleration $a$ relative to a detector has been given in [7], Section 3.3, by,

$$\Psi(\Delta E, a) = (\text{const.}) \sum_{\Delta E} \Delta E \frac{|\langle E | m(0) | E_0 \rangle|^2}{e^{2\mathbf{p}\Delta E/a} - 1}, \qquad (2)$$

in which the energy $\Delta E = E - E_0$ of a radiative transition depends on properties of the Minkowski vacuum. From this expression, the effective temperature $T$ for particle creation (most directly, photon creation) is related [7] to acceleration by,

$$T = \frac{a}{2\mathbf{p} k_B}. \qquad (3)$$

Clearly, (1) is in the same form as any one term of (2), allowing for a numerator of the latter on uniform wavelength scale. Using (3), then, at any fixed frequency $\mathbf{n}$ in Eq. (1) we may write,

$$I_\mathbf{n}(a) = \frac{\mathbf{k}_1}{e^{-\mathbf{k}_2/a} - 1}, \qquad (4)$$

for acceleration $a$, and $\mathbf{k}_1$ and $\mathbf{k}_2$ frequency-dependent constants. The constants $\mathbf{k}_1$ and $\mathbf{k}_2$ are independent, containing different powers of $\mathbf{n}$, but are not orthogonal.

To express this distribution as a sum over any domain of accelerations $\{a\}$, we must integrate (4). Changing variables and integrating by parts,

$$I_\mathbf{n}\{a\} = \int_{\{a\}} da \frac{\mathbf{k}_1}{e^{-\mathbf{k}_2/a} - 1} = \int_{\{a^{-1}\}} dx \frac{-\mathbf{k}_1 x^{-2}}{e^{-\mathbf{k}_2 x} - 1} = a\frac{-\mathbf{k}_1}{e^{-\mathbf{k}_2/a} - 1} + \mathbf{k}_2 \int_{\{a^{-1}\}} dx \frac{-\mathbf{k}_1 x^{-1}}{(e^{-\mathbf{k}_2 x} - 1)^2}. \qquad (5)$$

Repeated integrations by parts do not yield a simple convergent series; and, inspection reveals that, because of the lack of a $k_2$ coefficient, the first term on the right of (5) never can be cancelled.  Therefore, (5) never can be of the form of (4) and so can not be Planckian.  Also, thermodynamically, a spatially distinct set of different accelerations (temperatures) could in principle be used by a heat engine to extract free energy:  So, in general, the spectrum from particles distributed in a continuum of extreme accelerations (temperatures) will not be Planckian.  This holds whether the particle creation be assumed to follow Bose-Einstein statistics (*e. g.*, photons as above), Fermi-Dirac statistics (*e. g.*, neutrinos), or Boltzmann statistics.   Some puzzling speculations about loss of information into black holes may be resolved by pursuing this direction of analysis.

To examine further our own limiting case of extremely high acceleration, we differentiate (4) with respect to $a$,

$$I_n'(a) = \frac{-k_1 k_2 \cdot e^{-k_2/a}}{a^2 (e^{-k_2/a} - 1)^2} = \frac{-k_1 k_2}{a^2 (e^{-k_2/a} + e^{k_2/a} - 2)} . \tag{6}$$

We find that as $a \to \infty$, $I_n'(a) \to -k_1/2 = -p\hbar n^3/c^2$, which is a constant function of acceleration.  But, in general, there is no reason to assume all extremely high values of acceleration will be equal; so, the radiation in any detector not also extremely accelerated will be a linear sum of contributions from some range of different accelerations, $\{a\}$.  From (6) and immediately after, such a detector will report a spectral density slope determined within an additive constant according to,

$$\lim_{a \to \infty} \frac{\partial}{\partial a} \Psi(n, \{a\}) = -\lim_{a \to \infty} \int_{\{a\}} da \left( \frac{p\hbar n^3}{c^2} a + \text{const} \right). \tag{7}$$

Because the law of change of $\{a\}$ with increasing $a$ is not specified, we can say no more about (7) except that the spectral derivative will not approach that of a thermal distribution.

We conclude that in the sense of (5), and of the limit in (7), a black hole will not emit black-body radiation.  This, because of the variation of acceleration with radial distance near the event horizon and the finite probability of pair-creation at any moment at more than a single value of the radial distance.  For a given set of radiative accelerations $\{a\}$, the narrower the range of radial distances over which radiation from a black hole occurred, the greater would be the departure from a Planckian spectrum, in that the greater would be the difficulty of preventing detector response to all but a narrow, approximately Planckian, range of such accelerations.  This, of course, is familiar to astronomers, who distinguish, for example, the emission of the Sun's photosphere at $\sim 10^{3.8}$ K from that of its corona at $\sim 10^6$ K.

# Existence Proof

A virtual particle can exist only in a time-like interval; otherwise, its creation and annihilation points would be separated by a space-like interval, and, therefore, it would have to do work to annihilate. For example, Čerenkov radiation is produced when a change of medium puts some of the energy of a highly localized Coulomb potential into an effectively space-like interval because of speed in the new medium. Real photons are created because the Coulomb virtual photons find themselves having to do work locally instead of being exchanged electromagnetically. This, however, first requires a speed above that of light in the new medium. No such speed can be attained in vacuum.

For our purposes, we invoke special relativity to the extent that we consider acceleration, velocity, time, and displacement to be defined in the proper frame of the object being accelerated. For an existence proof, we require only that measurement of the proper variables in the accelerated object's frame be monotonically related to their measurement the lab frame: An increase in the proper frame maps to an increase in the lab frame, however scaled, and likewise a decrease. In this way, we avoid calculation of Lorentz factors entirely, as well as all considerations of the space-time metric.

We consider the vacuum as a frame-independent entity defined solely by energy. Because momentum is not frame-independent, we expect to be able to use the vacuum as an operator to separate energy from momentum. Particles can not be so operated upon, so we expect to be able to quantify particle creation from the vacuum, in some sense, by examining both the energy and the momentum uncertainty of particles created from vacuum.

Consider two statements of Heisenberg's uncertainty principle for a free particle:

$$\Delta E \cdot \Delta t \geq \frac{\hbar}{2}; \text{ and,} \tag{8}$$

$$\Delta p \cdot \Delta x \geq \frac{\hbar}{2}. \tag{9}$$

A massless particle moves at the speed of light and with energy $E = cp = h\mathbf{n}$. Acceleration is meaningful only for a massive particle; so, from (9), for a particle of mass $m$, the momentum $p = mv$; and,

$$\Delta p \cdot \Delta x \geq \frac{\hbar}{2} \quad \Rightarrow \quad \Delta(mv) \cdot \Delta x \geq \frac{\hbar}{2}. \tag{10}$$

So, the $p$ uncertainty complementary to that in $x$ may be seen as distributed between $m$ and $v$. For a familiar particle of rest mass $m$, such as an electron, an experiment may be designed so as to take advantage of knowledge of the mass value. Therefore, from here on, we assume a measurement of $\Delta(mv)$ in (10) such that the uncertainty in the $v$ (speed) factor will be much greater than that in the $m$ factor. We thus assume that,

$$\frac{d}{d(\Delta x)}m \ll \frac{d}{d(\Delta x)}v \quad\Rightarrow\quad \frac{d(mv)}{d(\Delta x)} = v\frac{dm}{d(\Delta x)} + m\frac{dv}{d(\Delta x)} \cong m\frac{dv}{d(\Delta x)}. \tag{11}$$

We write this relation mnemonically as $\Delta(mv) \equiv \boldsymbol{dn} \cdot \Delta v$. So, $|\boldsymbol{dn}|$ will be relatively small when compared with $|\Delta v|$; and, during acceleration we expect $|\boldsymbol{dn}|$ (the uncertainty in the rest mass) to remain constant or perhaps to change in the same direction as $|\Delta v|$.

We wish to show that under high acceleration of our particle, vacuum-acceleration radiation always will result. Thus, the virtual interval containing the accelerated particle and any new one(s) created from vacuum always may be made to shrink to make any new interval become space-like; and, so, at some acceleration, a shower of real vacuum radiation (particles) will occur.

To do this, we wish to prove an expression that, for some variable $V$ representing either energy $E$ or momentum $p$, as acceleration $a \to \infty$, $\Delta x/\Delta V$ must shrink to 0. If the ratio of $\Delta x$ to our $\Delta V$ did shrink to 0, eventually every interval of virtual particles containing $V$ would become space-like, and we would have emission of real vacuum particle(s) solely because of acceleration of our massive particle.

Let's assume that the proper duration of existence of a virtual particle created from the vacuum during acceleration was reasonably precisely measured; so, in (8), we will have a fairly small uncertainty $\Delta t$ and a complementarily large $\Delta E$, in the lab frame. We also assume we can measure a propagation interval $\Delta x$ with some precision in the lab frame. In this way, acceleration may be defined reasonably well. From (10),

$$(\boldsymbol{dn} \cdot \Delta v) \cdot \Delta x \geq \frac{\hbar}{2}. \tag{12}$$

During acceleration, we will have the lab frame speed in the direction of acceleration changing so that $|v| \to c$. By definition of the derivative,

$$a = \frac{dv}{dt} = \lim_{\Delta t \to 0} \frac{\Delta v}{\Delta t}, \tag{13}$$

in which the deltas here are not defined as uncertainties but rather as signed differences: $\Delta t \equiv t_1 - t_0 > 0$ and $\Delta v \equiv v(t_1) - v(t_0)$. Clearly, if we allow acceleration to increase $v$ by some large amount, the sign of $\Delta v$ in (12), as an uncertainty, will be correct for (13), as a difference. So, we have made the transition from differences to uncertainties, as has been done in [6] and elsewhere. Using the standard deviation *sd* as a measure of uncertainty which preserves units, we may write a quotient of Heisenberg uncertainties so that,

$$\frac{\Delta x}{\Delta t} = \frac{sd(x)}{sd(t)}. \tag{14}$$

Now we only need assume that the *sd* of *x* will increase with increased *x*, as measured during a given *t*. In that case, letting $v \to c$ means that both sides of (14) also increase toward *c*. By the mean value theorem, we therefore conclude for a quotient of uncertainties that,

$$\frac{\Delta v}{\Delta t} = \frac{sd(v)}{sd(t)} \quad \Rightarrow \quad \lim_{a \to \infty} \frac{\Delta v}{\Delta t} = \infty, \tag{15}$$

recalling that there is no relativistic limit to acceleration.

Returning to (12), and dividing through by $\Delta t$,

$$\frac{\Delta v}{\Delta t} \cdot dn \cdot \Delta x \geq \frac{\hbar}{2\Delta t}; \text{ so,} \tag{16}$$

$$\frac{2dn\Delta t \Delta x}{\hbar} \geq \frac{1}{\Delta v/\Delta t}. \tag{17}$$

Using (15) to reexpress $\Delta t$,

$$\Delta E \cdot \Delta t \geq \frac{\hbar}{2} \quad \Rightarrow \quad \Delta t \geq \frac{\hbar}{2\Delta E}. \tag{18}$$

We may use the right side of (18) as an equality if we take $\Delta E$ to be a lower bound; we write this constrained $\Delta E$ as $\underline{\Delta E}$. With this, substituting the rightmost expression in (18) for $\Delta t$ on the left in (17),

$$\frac{2dn \frac{\hbar}{2\underline{\Delta E}} \Delta x}{\hbar} \geq \frac{1}{\Delta v/\Delta t} \quad \Rightarrow \quad \frac{dn\Delta x}{\underline{\Delta E}} \geq \frac{1}{\Delta v/\Delta t}; \tag{19}$$

and, finally,

$$\frac{\Delta u}{\Delta t} \geq \frac{\underline{\Delta E}}{dn\Delta x} \quad \Rightarrow \quad a \geq \left(\frac{1}{dn}\right)\frac{\underline{\Delta E}}{\Delta x}. \tag{20}$$

In (20), simply letting acceleration $a \to 0$, we must have $\Delta x$ very large relative to $\underline{\Delta E}$. However, in any one measurement apparatus, if we wish to let $a \to \infty$ beginning from some low value of $a$ not far from 0, we find that we must increase $\Delta E$ relative to $\Delta x$. Therefore, as $a \to \infty$ we must have,

$$\left(\frac{1}{d n}\right)\frac{\Delta E}{\Delta x} \to \infty; \qquad (21)$$

and, because of the postulated relatively low rate of change of $dn$, it must be that $\Delta x$ will shrink toward 0 relative to $\Delta E$ in the limit. So,

$$\lim_{a \to \infty} \frac{\Delta x}{\Delta E} = 0. \qquad (22)$$

Therefore, there must be some acceleration at which any virtual particle from the vacuum will be separated in the lab frame from its proper locus of creation by a space-like interval, thus becoming real.   Q. E. D.

## Implications

We note that as acceleration is increased, because a given interval will be space-like earliest for particles travelling on the light cone, the earliest vacuum-acceleration particles to be created would be expected to be photons or other massless particles.

However, although the ratio in (22) will approach 0, the value of $\Delta x$ in general will increase with acceleration; $\Delta E$ will increase also, and at a greater rate. Because at particle-creating accelerations Planckian energy will be very uncertain over any spatial region discriminable by a detector in the original rest frame, the unavoidably integrated spectrum of the particle(s) detected, as in (5) above, will not be thermal.   Because we have not specified the force or reaction force of the acceleration, we can not here draw a more specific conclusion about this spectrum.

As mentioned in the Introduction, the existence of a vacuum radiation threshold for acceleration implies the existence of a lower limit on the mass of a black hole, however cataclysmically formed.   It also leads to new perspectives on elementary particles, as in the following:

Consider the classical Coulomb force between the massive elementary particles (quarks) in a hadron, such as a meson, proton, or neutron: Why don't these latter particles evaporate because of vacuum acceleration radiation by the supposedly point-like quarks?   After all, the Coulomb acceleration on a $b - \bar{b}$ system of point-like quarks with occasional diameter 1 fm occasionally must be at least some $10^{65}\,\text{m s}^{-2}$. Our proof does not restrict itself to the electromagnetic quanta of de Broglie's orbits; so, one might expect radiation of something.   Hosoya [8] has calculated a thermal spectrum in this context for hadronic internal interactions. Compare the $b - \bar{b}$ acceleration above with the acceleration from gravity of a black hole of 10 solar masses: The horizon radius would be, $r = 2G_N M/c^2 \cong 4 \cdot 10^4$ m; so, the acceleration at the horizon only would be, $a = c^2/2r \cong 10^{12}$ m s$^{-2}$.

Why no radiative evaporation of hadrons? The usual answer would be to point out that the question should be directed to the quantum realm, and that uncertainty in $\Delta x$ or $\Delta p$ definitely would apply. But, an answer from the present work would be to put the uncertainty in a different place: In the CKM quark-mixing matrix, the off-diagonal terms imply that the masses of the quarks must be to some extent indefinite; this means that $dh$ in Eq. (21) would be larger than otherwise--large enough, perhaps, to prevent Coulomb acceleration from reaching a radiative threshold. So, the strong force, by making the elementary masses a little indefinite, might be seen as precluding destructive values of acceleration.

If this speculation were to be pursued a little further, one might contemplate the possibility of extracting energy from the light elements by devising a way to cause the CKM matrix to diagonalize.

# Acknowledgements

The author thanks Mario Rabinowitz for comments on a draft.